\documentclass[twocolumn,tighten]{aastex62}
\usepackage{graphicx}
\usepackage{bm}

\received{March 11, 2019}
\revised{September 17, 2019}
\accepted{September 18, 2019}

\shorttitle{Enrichment of Sr}
\shortauthors{Hirai et al.}

\begin{document}

\title{Enrichment of Strontium in Dwarf Galaxies}

\correspondingauthor{Yutaka Hirai}
\email{yutaka.hirai@riken.jp}

\author[0000-0002-5661-033X]{Yutaka Hirai}
\affiliation{RIKEN Center for Computational Science, 7-1-26 Minatojima-minami-machi, Chuo-ku,
Kobe, Hyogo 650-0047, Japan}
\author[0000-0002-4759-7794]{Shinya Wanajo}
\affiliation{Max-Planck-Institut f\"ur Gravitationsphysik (Albert-Einstein-Institut), Am M\"uhlenberg 1, D-14476 Potsdam-Golm, Germany}
\affiliation{RIKEN iTHEMS Research Group, 2-1 Hirosawa, Wako, Saitama 351-0198, Japan}
\affiliation{Department of Engineering and Applied Sciences, Faculty of Science and Technology, Sophia University, 7-1 Kioicho, Chiyoda-ku, Tokyo 102-8554, Japan}
\author[0000-0001-8226-4592]{Takayuki R. Saitoh}
\affiliation{Department of Planetology, Graduate School of Science, Kobe University, 1-1, Rokkodai-cho, Nada-ku, Kobe, Hyogo 657-8501, Japan}
\affiliation{Earth-Life Science Institute, Tokyo Institute of Technology, 2-12-1 Ookayama, Meguro-ku, Tokyo 152-8551, Japan}

\begin{abstract}Light trans-iron elements such as Sr serve as the key to understanding the astrophysical sites of heavy elements. Spectroscopic studies of metal-poor stars have revealed large star-to-star scatters in the ratios of [Sr/Ba], which indicates that there are multiple sites for the production of Sr. Here we present the enrichment history of Sr by a series of the $N$-body/smoothed particle hydrodynamics simulations of a dwarf galaxy with a stellar mass of 3 $\times$ 10$^{6}$ $M_{\sun}$. We show that binary neutron star mergers (NSMs) and asymptotic giant branch (AGB) stars contribute to the enrichment of Sr in the metallicity ranges [Fe/H] $\gtrsim$ $-$3 and [Fe/H] $\gtrsim$ $-$1, respectively. It appears insufficient, however, to explain the overall observational trends of Sr by considering only these sites. We find that the models including electron-capture supernovae (ECSNe) and rotating massive stars (RMSs){, in addition to NSMs and AGBs}, reasonably reproduce the enrichment histories of Sr in dwarf galaxies. The contributions of both ECSNe and NSMs make scatters {of $\approx$ 0.2 dex} in [Sr/Fe], [Sr/Ba], and [Sr/Zn] as can be seen for observed stars in the metallicity range [Fe/H] $<$ $-2$. We also find that the mass range of ECSN progenitors should be substantially smaller than $1\, M_\odot$ (e.g., 0.1--$0.2\, M_\odot$) to avoid over-prediction of [Sr/Ba] and [Sr/Zn] ratios. Our results demonstrate that NSMs, AGBs, ECSNe, and RMSs all play roles in the enrichment histories of Local Group dwarf galaxies, although more observational data are required to disentangle the relative contributions of these sources.
\end{abstract}


\section{Introduction} \label{sec:intro}
\setcounter{footnote}{6}
The astrophysical sites of heavy elements are one of the open questions in astronomy over 60 years \citep{1957RvMP...29..547B, 1957PASP...69..201C}. The elements heavier than Fe are synthesized by neutron-capture processes. These are divided into the $r$- and $s$-processes according to the timescales of neutron capture be faster and slower than those of competing $\beta$-decay on relevant isotopes, respectively. While the astrophysical sites of the $s$-process are relatively well understood \citep[e.g.,][]{2011RvMP...83..157K}, those of the $r$-process remains a mystery \citep[e.g.,][]{2017ARNPS..67..253T, 2019arXiv190101410C}.

Understanding of the $r$-process has been significantly improved in recent years. Binary neutron star mergers (NSMs) are taken to be the most promising astrophysical site of the $r$-process since the discovery of the gravitational waves from an NSM, GW170817, measured by the Advanced LIGO and Virgo \citep{2017PhRvL.119p1101A}. Follow-up observations suggest that the ejecta containing a small fraction of lanthanides ($\lesssim 0.01$ in mass) can explain the light curves of the electromagnetic counterpart of GW170817 \citep[e.g.,][]{2017ApJ...848L..19C, 2017ApJ...848L..17C, 2017Sci...358.1570D, 2017Natur.551...67P, 2017PASJ...69..102T, 2017ApJ...851L..21V, 2018ApJ...868...65W}. This is in accordance with the recent nucleosynthetic studies based on the numerical simulations of NSMs. These studies suggest that the early dynamical ejecta contain all $r$-process nuclides including lanthanides and heavier ones \citep{2014ApJ...789L..39W, 2015PhRvD..91f4059S, 2016PhRvD..93l4046S, 2015MNRAS.452.3894G, 2016MNRAS.460.3255R, 2018PhRvD..98j4028P} and the secular disk outflows the lighter elements with few lanthanides \citep{2018PhRvD..98j4028P, 2017MNRAS.472..904L, 2018ApJ...860...64F}. 

Spectroscopic studies of metal-poor stars in the Milky Way (MW) halo have provided us with numerous clues to understanding the origin of heavy elements. The abundance distributions of highly $r$-process-enhanced stars closely match the solar-system $r$-process pattern in the atomic number ($Z$) range, 56 $\leq$ $Z$ $<$ 80 \citep[e.g.,][]{2008ARA&A..46..241S}. This result implies the presence of a single, robust astrophysical site at least for these $r$-process elements. On the other hand, the elements with $Z$ $<$ 56, including the lightest neutron-capture (or a light trans-iron) species, Sr-Y-Zr, show a factor of several of variation compared to the scaled solar $r$-process pattern. This fact indicates that the nucleosynthetic conditions for light trans-iron elements such as Sr are not as robust as those for the heavy $r$-process elements.

\begin{figure}[htbp]
\epsscale{1.3}
\plotone{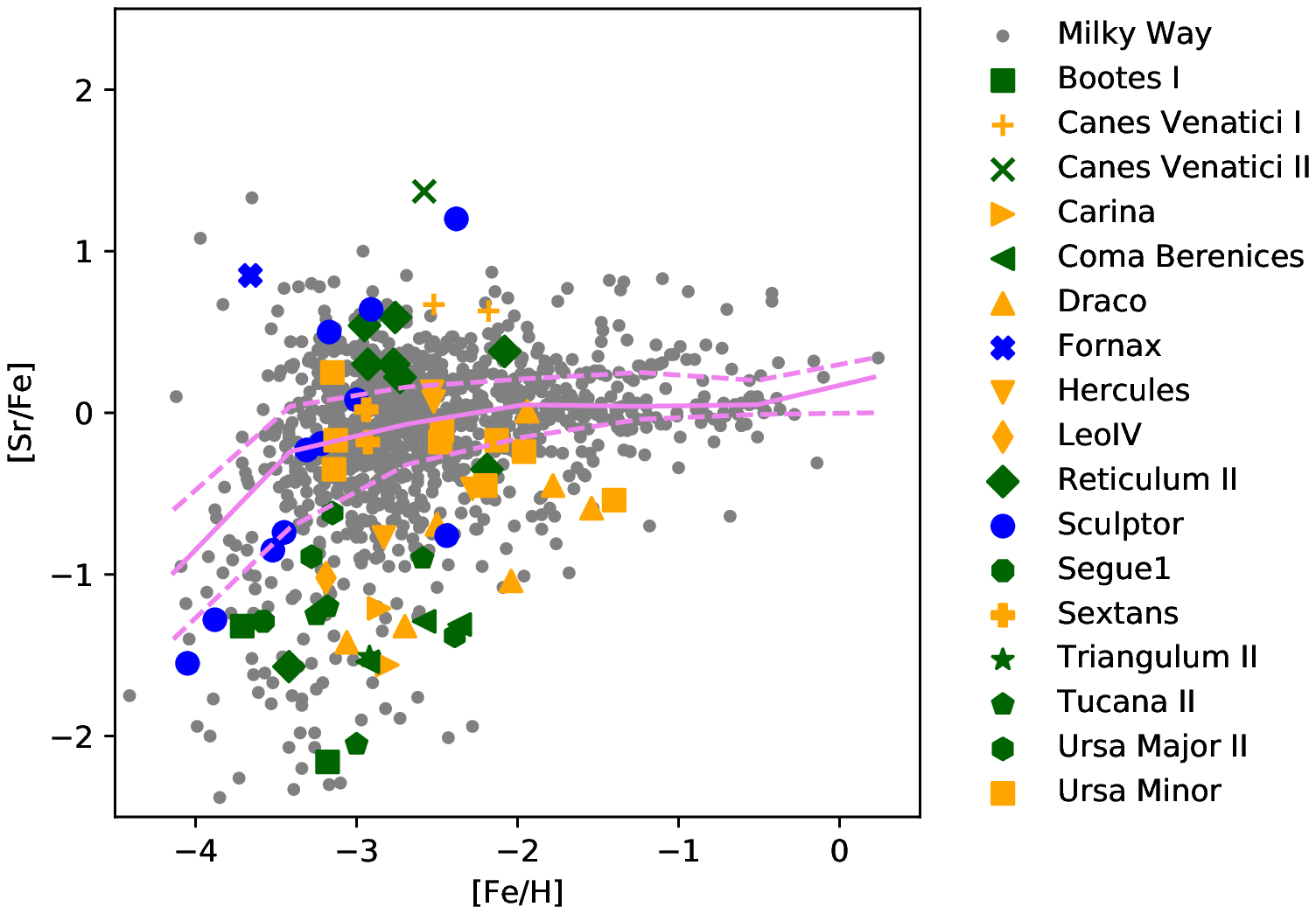}
\caption{Measured [Sr/Fe] ratios as a function of [Fe/H]. Green symbols denote stars in the dwarf galaxies with the stellar masses $M_{*}$ $<$ 10$^5$ $M_{\sun}$: Bo\"otes I \citep[squares,][]{2010ApJ...711..350N, 2014A&A...562A.146I}, Canes Venatici II \citep[crosses,][]{2016A&A...588A...7F}, Coma Berenices \citep[left triangles][]{2010ApJ...708..560F}, Reticulum II \citep[diamonds,][]{2016AJ....151...82R, 2016ApJ...830...93J, 2018ApJ...856..138J}, Segue 1 \citep[octagons,][]{2010ApJ...722L.104N, 2014ApJ...786...74F}, Triangulum II \citep[stars,][]{2017ApJ...838...83K}, Tucana II \citep[pentagons,][]{2018ApJ...857...74C}, and Ursa Major II \citep[hexagons,][]{2010ApJ...708..560F}. Orange symbols represent those with 10$^5$ $M_{\sun}$ $\leq$ $M_{*}$ $<$ 10$^6$ $M_{\sun}$: Canes Venatici I \citep[pluses,][]{2016A&A...588A...7F}, Carina \citep[right triangles][]{2012ApJ...751..102V}, Draco \citep[triangles,][]{2001ApJ...548..592S, 2004ApJ...612..447F, 2009ApJ...701.1053C,2015PASJ...67L...3T}, Hercules \citep[inverted triangles][]{2016A&A...588A...7F}, Leo IV \citep[thin diamonds,][]{2010ApJ...716..446S}, Sextans \citep[filled pluses,][]{2010A&A...524A..58T}, and Ursa Minor \citep[hexagons,][]{2001ApJ...548..592S, 2004PASJ...56.1041S,2010ApJ...719..931C, 2012AJ....144..168K}. Blue symbols depict those with 10$^6$ $M_{\sun}$ $\leq$ $M_{*}$: Fornax \citep[filled crosses,][]{2003AJ....125..684S, 2010A&A...523A..17L, 2010A&A...524A..58T, 2014A&A...572A..88L} and Sculptor \citep[circles,][]{2003AJ....125..684S,2005AJ....129.1428G, 2012AJ....144..168K, 2015A&A...583A..67J,2015ApJ...802...93S, 2017A&A...608A..89M}.  Gray circles are stars in the MW. Magenta solid and dashed curves indicate the median and the first and third quantiles, respectively. All data are compiled using the SAGA database \citep{2008PASJ...60.1159S, 2011MNRAS.412..843S, 2017PASJ...69...76S, 2013MNRAS.436.1362Y}, in which carbon-enhanced stars (whose abundances may be affected by binary mass transfer) are excluded. \label{SrFeobs}}
\end{figure}

Studies of metal-poor stars in the MW halo show large star-to-star scatters in the ratio of Sr-Y-Zr to Fe \footnote{Original references are  \citet{1995AJ....109.2757M, 1998ApJ...506..892R,2000AJ....120.1014P,2000ApJ...544..302B, 2001A&A...370..951M,  2002ApJ...572..861C, 2002AJ....124..481C, 2002ApJ...579..616J, 2003AJ....125..875L, 2003ApJ...592..906I, 2004ApJ...607..474H, 2007ApJ...666.1189H,2011ApJ...730...77H, 2005A&A...439..129B, 2005ApJ...632..611A, 2007ApJ...660..747A, 2009ApJ...698.1803A, 2010ApJ...723L.201A, 2013AJ....145...13A, 2014Sci...345..912A, 2006AJ....132...85P, 2006A&A...455.1059M, 2006A&A...451..651J,  2007A&A...476..935F, 2007ApJ...667.1185L, 2008ApJ...681.1524L, 2008ApJ...672..320C, 2013ApJ...778...56C, 2008A&A...478..529M, 2014A&A...569A..43M, 2009A&A...501..519B,   2011ApJ...742...54H, 2011ApJ...743..107R, 2012A&A...542A..87B, 2012ApJS..203...27R, 2014AJ....147..136R, 2016ApJ...821...37R, 2012A&A...545A..31H, 2017A&A...598A..54H, 2013A&A...558A..36C, 2013ApJ...771...67I, 2013ApJ...762...26Y,  2014A&A...565A..93S,  2014A&A...571A..40S, 2014A&A...571A..62C, 2015A&A...582A..74S, 2015ApJ...798..110L, 2015ApJ...807..171J, 2015ApJ...807..173H, 2016MNRAS.458.2648S, 2017ApJ...837....8A}.}, e.g., [Sr/Fe]\footnote{[A/B] = $\log_{10}({N_{\mathrm{A}}}/{N_{\mathrm{B}}})-\log_{10}({N_{\mathrm{A}}}/{N_{\mathrm{B}}})_{\odot}$, where $N_{\mathrm{A}}$ and $N_{\mathrm{B}}$ are the numbers of the elements A and B, respectively.} \citep[Figure \ref{SrFeobs}, see also][]{1996ApJ...471..254R, 1998AJ....115.1640M, 2004ApJ...607..474H, 2015ARA&A..53..631F}, a similar trend to those in [Ba/Fe] and [Eu/Fe]. However, the ratios of these elements to heavy neutron-capture elements (e.g., [Sr/Ba]) also show scatters \citep[e.g.,][]{1998AJ....115.1640M, 2002ApJ...579..616J, 2013AJ....145...13A, 2015ApJ...807..171J}.
Recent surveys by the R-Process Alliance collaboration have found that 39 out of 233 stars exhibit [Sr/Ba] $>$ 0.5 \citep{2018ApJ...858...92H, 2018ApJ...868..110S}, which is substantially greater than the solar $r$-process value: [Sr/Ba] = $-$0.13 \citep{2004ApJ...617.1091S}. \citet{2007A&A...476..935F} found a tight anti-correlation of [Sr/Ba] ratios as a function of [Ba/H] (i.e., the enhancement of $r$-process elements).  These studies suggest the presence of a second astrophysical source for the light trans-iron elements, such as Sr, at low metallicity \citep{1998AJ....115.1640M}.
 
The Local Group (LG) dwarf galaxies also indicate a second source of Sr. In the dwarf galaxy Draco, only the lowest metallicity star shows the pure $r$-process ratio of [Sr/Ba] \citep{2009ApJ...701.1053C}. The other stars show neither signs of the pure $r$-process nor the pure $s$-process. The dwarf galaxies Canes Venatici I and II have stars with significantly enhanced [Sr/Ba] ratios \citep{2016A&A...588A...7F}. These results cannot be explained by a single source that leads to nearly a constant ratio of [Sr/Ba].

The astrophysical site of the second source of Sr at low metallicity has not yet been clarified. Core-collapse supernovae (CCSNe) from low mass progenitors may contribute to the enrichment of light trans-iron elements. Stars at the low-mass end ($\sim$ 8--10 $M_{\sun}$) develop O-Ne-Mg cores and end their lives either as white dwarfs or electron-capture supernovae,  ECSNe\footnote{ECSNe are a subset of CCSNe but with collapsing O-Ne-Mg cores. In this study, we do not consider ``thermonuclear" ECSNe \citep{2019A&A...622A..74J,2019arXiv190802236J} that can be the source of $^{48}$Ca, $^{50}$Ti, and $^{54}$Cr but not of Sr-Y-Zr.}\citep[e.g.,][]{1980PASJ...32..303M, 1982Natur.299..803N, 1984A&A...133..175H, 1984ApJ...277..791N, 1987ApJ...322..206N, 1987ApJ...318..307M}. The mass range that leads to the ECSN channel is highly sensitive to the stellar evolution models adopted \citep{2008ApJ...675..614P, 2017PASA...34...56D} as well as to the binarity of the systems \citep{2017ApJ...850..197P, 2018A&A...614A..99S}. \citet{2011ApJ...726L..15W, 2018ApJ...852...40W} have shown that ECSNe and low mass iron-core CCSNe\footnote{In \citet{2018ApJ...852...40W}, the $9.6\, M_\odot$ CCSN model with iron-core exhibits almost the same nucleosynthetic outcomes with those of an ECSN model. Hereafter, we intend that the term ``ECSNe" intrinsically includes such iron-core CCSNe at the low-mass end.} synthesize light trans-iron elements from Zn to Sr-Y-Zr, but the ejecta are not neutron-rich enough to produce heavy neutron-capture elements. However, light $r$-process elements up to Pd and Ag can be made in ECSNe if the ejecta are slightly more neutron-rich than those predicted in their original model \citep{2011ApJ...726L..15W}. Such a nucleosynthetic signature is called a ``weak $r$-process" \citep{2006NuPhA.777..676W} or a ``limited $r$-process" \citep{2018ARNPS..68..237F}, which is suggested to be the source of the descending abundance trends of neutron-capture elements in stars with high [Sr/Ba] ratios \citep{2006ApJ...643.1180H, 2017ApJ...837....8A}.  

Several studies also have pointed out that the weak $s$-process in rotating massive stars (RMSs) may contribute to the enrichment of light trans-iron elements \citep[e.g.,][]{2011Natur.472..454C,2013A&A...553A..51C, 2014A&A...565A..51C, 2017ApJ...835...97B, 2018MNRAS.476.3432P, 2018ApJS..237...13L, 2018A&A...618A.133C, 2019MNRAS.tmp.2143R}. Recently, \citet{2018MNRAS.476.3432P} have demonstrated that RMSs can be a predominant source of Sr at [Fe/H] $> -2.5$ by using a one-zone model of Galactic chemical evolution with the yields of \citet{2018ApJS..237...13L}. The main $s$-process in asymptotic giant branch (AGB) stars also eject these elements \citep[e.g.,][]{2009ApJ...696..797C, 2011ApJS..197...17C, 2015ApJS..219...40C}, which becomes important only at high metallicity \citep{2018MNRAS.476.3432P}.

Contributions of ECSNe to the galactic chemical evolution of light trans-iron elements such as Sr are not well understood. \citet{2014A&A...565A..51C} implemented the empirical yields of ECSNe in their inhomogeneous chemical evolution models. They, however, assume the production of Ba, which is unlikely according to the recent nucleosynthesis calculations using the hydrodynamic simulations of ECSNe \citep{2011ApJ...726L..15W, 2018ApJ...852...40W}. \citet{2018ApJ...865...87O} also included the empirical yields of ECSNe in their chemical evolution model but did not explore the uncertainties in the progenitor mass range resulting from different stellar evolution models. \citet{2018ApJ...855...63H} performed chemo-dynamical simulations of dwarf galaxies with ECSNe only focusing on the enrichment of Zn. 

Recent nucleosynthesis studies of CCSNe suggest that the proton-rich material dominates in the  neutrino-driven ejecta with sub-dominant production of Sr \citep{2015ApJ...808..188P, 2018ApJ...852...40W}. It is unfeasible, however, to follow a long-term evolution of neutrino-driven wind (over $\sim 1$~s) in current multi-dimensional computations. Although the bulk of neutrino-driven ejecta are expected to be proton-rich, only a slight neutron-richness leads to substantial production of Sr \citep{2013ApJ...770L..22W}. {A possible contribution from the neutrino-driven ejecta of CCSNe to the enrichment history of Sr is not excluded.}

This paper aims to clarify the contributions of {possible sourses of Sr (NSMs, AGBs, ECSNe, and RMSs)} to galactic chemical evolution. {We do not consider the contribution from the neutrino-driven wind of CCSNe because of the absence of realistic nucleosynthetic yields currently available.} In this study, we focus on the enrichment of Sr in dwarf galaxies, although the number of stars with the measurements of Sr is still limited and much smaller than that in the MW (Figure \ref{SrFeobs}). This is because dwarf galaxies are the ideal objects to clarify the roles of nucleosynthetic sites because of their simple structures and evolutionary histories \citep[e.g.,][]{2015ApJ...807..154B}. Due to the lack of observed data in dwarf galaxies, we will statistically compare our result also with the data of the MW in this study. This may be justified by assuming that dwarf galaxies are in part the leftovers of building blocks that have made the MW \citep[e.g.,][]{2005ApJ...635..931B, 2015ApJ...804L..35I, 2018ApJ...865...87O}.  {We perform} high-resolution simulations, including the detailed models of nucleosynthetic sites and metal mixing by focusing on dwarf galaxies. 

This paper is organized as follows. Section \ref{sec:method} describes methods and models adopted in this study. Section \ref{sec:results} shows the roles of NSMs, AGBs, ECSNe, and RMSs on the enrichment of Sr in dwarf galaxies. Section \ref{sec:dis} discusses the astrophysical sites of Sr, along with other heavy elements such as Zn and Ba. In Section \ref{sec:con}, we present conclusions of this study.

\section{Method} \label{sec:method}
\subsection{Code}
In this study, we perform a series of $N$-body/smoothed particle hydrodynamics (SPH) simulations using \textsc{asura} \citep{2008PASJ...60..667S, 2009PASJ...61..481S}. Details of the code are described in \citet{2015ApJ...814...41H, 2017MNRAS.466.2474H}. Here we briefly describe our code. Gravity is computed with the tree method \citep{1986Natur.324..446B} parallelized following \citet{2004PASJ...56..521M}. We adopt the density-independent formulation of SPH (DISPH) to handle the fluid instabilities in a contact discontinuity correctly \citep{2013ApJ...768...44S, 2016ApJ...823..144S}. The FAST scheme is adopted to reduce the computation cost of the strong shock regions \citep{2010PASJ...62..301S}. We keep the difference of the timestep in neighbor particles small enough using the timestep limiter \citep{2009ApJ...697L..99S}. To compute radiative cooling, we adopt the metallicity-dependent cooling/heating functions generated by the version 13.05 of \textsc{Cloudy} from 10 to 10$^9$ K \citep{1998PASP..110..761F, 2013RMxAA..49..137F, 2017RMxAA..53..385F}. We also put the effect of self-shielding \citep{2013MNRAS.430.2427R} and the ultra-violet background radiation field \citep{2012ApJ...746..125H}.

We treat each star particle as a simple stellar population (SSP) assuming the initial mass function of \citet{2001MNRAS.322..231K} from 0.1 to 120 $M_{\sun}$. To convert gas particles to star particles, we set the threshold density and temperature as 100 cm$^{-3}$ and 1000 K, respectively, following \citet{2008PASJ...60..667S}. The number of Lyman-$\alpha$ photons from massive stars, which is used to compute the H$_{\rm{II}}$ regions, is evaluated using \textsc{p\'egase} \citep{1997A&A...326..950F}.

Stellar feedback is implemented using Chemical Evolution Library \citep[\textsc{celib},][]{2017AJ....153...85S}. Stars with 13--40 $M_{\sun}$ are assumed to explode as CCSNe. When a CCSN occurs, it distributes the thermal energy of 10$^{51}$ erg to surrounding gas particles. We take nucleosynthetic yields of \citet{2013ARA&A..51..457N} for Fe and Zn.

We adopt the same models of ECSNe in \citet{2018ApJ...855...63H}. The mass ranges of ECSNe are taken from stellar evolution calculations \citep{2015MNRAS.446.2599D}. At $10^{-4}$ $Z_{\sun}$, the mass range of progenitors is from 8.2 to 8.4 $M_{\sun}$. The mass range shifts to a more massive side at higher metallicity, while the mass window remains as narrow as $\Delta M = 0.1$--$0.2\, M_\odot$ \citep[see Table~1 in][]{2018ApJ...855...63H}. This mass range is affected by the uncertainties of stellar evolution such as mass-loss rates, third dredge up, and overshooting \citep{2017PASA...34...56D}. \citet{2018ApJ...855...63H} investigated the effects of uncertainties in the mass range of ECSNe on the enrichment of Zn. In Section \ref{ECSN}, we also present the result with a different mass range of ECSN progenitors.

The nucleosynthesis yields of ECSNe are taken from \citet{2018ApJ...852...40W}. In the e8.8 model of \citet{2018ApJ...852...40W}, a single ECSN ejects 7.9$\times$10$^{-5}$ $M_{\sun}$ of Sr. Although \citet{2018ApJ...852...40W} only have computed nucleosynthetic yields in the innermost ejecta, the abundances of heavy elements such as Sr are not affected by the outer H/He shell that does not add heavy elements. We regard Sr as representative of the light trans-iron elements as well as of the lightest $r$-process elements (Sr-Y-Zr) because of the largest numbers of measurements. As shown by \citet{2007A&A...476..935F}, [Y/Sr] ratios at low metallicity are nearly constant with small scatters, implying that both Y and Sr come from the same origin.

In this study, we also assume that five percent of progenitors between 20 and 40 $M_{\sun}$ explode as hypernovae (HNe) following \citet{2018ApJ...855...63H}. The fraction of HNe is taken from the estimated rates from the observations of long gamma-ray bursts \citep{2004ApJ...612.1044P, 2007ApJ...657L..73G}. HNe do not contribute to the enrichment of Sr but are one of the possible sites of Zn \citep[e.g.,][]{2002ApJ...565..385U, 2005ApJ...619..427U, 2006ApJ...653.1145K, 2007ApJ...660..516T}.

For type Ia supernovae (SNe Ia), we assume a power-law delay time distribution with the index of $-$1 based on \citet{2012PASA...29..447M} with the minimum delay time of 0.1 Gyr \citep[e.g.,][]{2008PASJ...60.1327T}. Nucleosynthetic yields of Fe from SNe Ia are taken from the N100 model of \citet{2013MNRAS.429.1156S}. 

We assume that low and intermediate-mass stars from 1 to 6 $M_{\sun}$ distribute elements by mass-loss during AGB phases. We adopt the yields of Sr and Ba in the \textsc{fruity} database \citep{2009ApJ...696..797C, 2011ApJS..197...17C, 2015ApJS..219...40C}. Details of the implementation are presented in \citet{2017AJ....153...85S}.

We also implement models of NSMs. We assume that 0.2 \% of stars with 8--20 $M_{\sun}$ cause NSMs. This rate is consistent with the estimation from the gravitational wave observations \citep{2017PhRvL.119p1101A} and observed binary pulsars \citep{2008LRR....11....8L}. The merger time distribution is assumed to be a power law with an index of $-$1 following the minimum merger time of 10 Myr \citep{2012ApJ...759...52D}. We assume that each NSM synthesizes 3.5 $\times$ 10$^{-4}$ $M_{\sun}${, 7.3 $\times$ 10$^{-5}$ $M_{\sun}$, and 1.6$\times$ 10$^{-5}$ $M_{\sun}$ of Sr, Ba, and Eu}, respectively \citep{2014ApJ...789L..39W}. {In this study, we do not consider a possible variation of NSM yields. This can lead to the under-prediction of scatters (to some extent) in the abundance ratios between these elements as described in Section \ref{sites}.} 

We ignored the energy deposition from NSMs because of their sufficiently lower rates than those of CCSNe. \citet{2017MNRAS.471.2088S} have confirmed that the effect of energy from NSMs is marginal in their simulations of ultra-faint dwarf galaxies (UFDs).

For metal mixing among SPH particles, we adopt the turbulence based metal diffusion model \citep{2010MNRAS.407.1581S, 2017AJ....153...85S, 2017ApJ...838L..23H}. Details of the implementation are shown in \citet{2017AJ....153...85S}. We take the scaling factor for the metal diffusion coefficient to be 0.01 following \citet{2017ApJ...838L..23H}.

To examine the role of RMSs, we also consider a model that adopts the yields of Sr and Ba in set F with stars of 13--40 $M_{\sun}$ \citep{2013ApJ...764...21C, 2018ApJS..237...13L}. The rotational velocity distributions of O- and early B-type stars are peaked at 175 kms$^{-1}$ and 100 kms$^{-1}$ in Large and Small Magellanic Clouds, respectively \citep{2008A&A...479..541H, 2009A&A...496..841H}. We thus take the yields computed with the rotational velocity of 150 kms$^{-1}$ as representative. As our purpose of including RMSs is to test their role at a qualitative level, we do not consider the metallicity dependence of rotational velocity distributions as that in \citet{2018MNRAS.476.3432P}.

\subsection{Initial conditions}
In this study, we adopt isolated dwarf galaxy models with the pseudo-isothermal profile \citep{2009A&A...501..189R, 2012A&A...538A..82R}. We take the same model adopted in \citet{2017ApJ...838L..23H}. The total mass of halo is 7.0 $\times$ 10$^8$ $M_{\sun}$. The truncation radius and core radius are set to be 8.9 and 1.3 kpc, respectively. We set the number of particles and gravitational softening length as 2.6 $\times$ $10^5$ and 7.8 pc, respectively. The masses of single dark matter and gas particles are 4.5 $\times$ 10$^{3}$ $M_{\sun}$ and 8.0 $\times$ 10$^{2}$ $M_{\sun}$, respectively.

\subsection{Models}
Table \ref{models} lists models adopted in this study. {All models include the contributions of NSMs and AGBs. Model A excludes the contributions of ECSNe and RMSs to clarify the roles of NSMs and AGBs on the enrichment of Sr (Section \ref{NSM}). Model B adds ECSNe with the mass range computed by \citet{2015MNRAS.446.2599D}.} Model {C} adopts a wider mass ranges of ECSN progenitors than that in model {B}, according to its uncertainty \citep[e.g.,][]{2017PASA...34...56D}. In model {C}, we also assume no ECSN event at $Z$ $\leq$ 10$^{-5}$ $Z_{\sun}$. This assumption is motivated by the studies of Population III star formation, which tends to predict top-heavy IMF \citep[e.g.,][]{2014ApJ...792...32S, 2015MNRAS.448..568H, 2016MNRAS.462.1307S}. The effect of mass ranges of ECSNe is discussed in Section \ref{ECSN}.  Model D adopts the yields of {ECSNe (as in model B) and} RMSs with the rotational velocity of 150 kms$^{-1}$. Section \ref{MS} discusses the contribution of RMSs on the enrichment of Sr.
\begin{deluxetable}{lllll}[htbp]
   \tabletypesize{\scriptsize}
   \tablecaption{List of models \label{models}} 
   \tablecolumns{2}
   \tablewidth{0pt}
   \tablehead{
      \colhead{Model}&
      Mass Ranges of ECSNe&
      RMSs&NSMs&AGBs}
      \startdata
      A&no&no&yes&yes\\
      B&\citet{2015MNRAS.446.2599D}&no&yes&yes\\
      C&8.2--9.2 $M_{\sun}$ for $Z$ $>$ $10^{-5}Z_{\sun}$&no&yes&yes\\
      D&\citet{2015MNRAS.446.2599D}&yes&yes&yes\\
      E&\citet{2015MNRAS.446.2599D}&no&no&yes\\
      F&\citet{2015MNRAS.446.2599D}&yes&no&yes\\
      \enddata
     \tablecomments{From left to right, columns show the name of models, metallicity ($Z$) dependent mass ranges of ECSNe, with (yes) or without (no) contribution of Sr and Ba from RMSs, NSMs, and AGBs. Models B$_1$ and B$_2$ in Appendix \ref{ap:NSMs} adopt the same parameters with model B but are computed with different random number seeds.}
\end{deluxetable}
\section{Results} \label{sec:results}
\subsection{Metallicity Distribution Functions}
Our model gives a metallicity distribution function (MDF) similar to the observed ones in the LG dwarf galaxies with similar masses. In model B, the stellar mass ($M_*$) and mean value of [Fe/H] ($\langle$[Fe/H]$\rangle$) are 3.2 $\times$ 10$^{6}$ $M_{\sun}$ and $-$1.45, respectively. These values are similar to those of the dwarf galaxy Sculptor ($M_*$ = 3.9 $\times$ 10$^{6}$ $M_{\sun}$ and $\langle$[Fe/H]$\rangle$ = $-$1.68) and Leo I ($M_*$ = 4.9 $\times$ 10$^{6}$ $M_{\sun}$ and $\langle$[Fe/H]$\rangle$ = $-$1.45)  \citep{2013ApJ...779..102K}.

\begin{figure}[htbp]
\epsscale{1.0}
\plotone{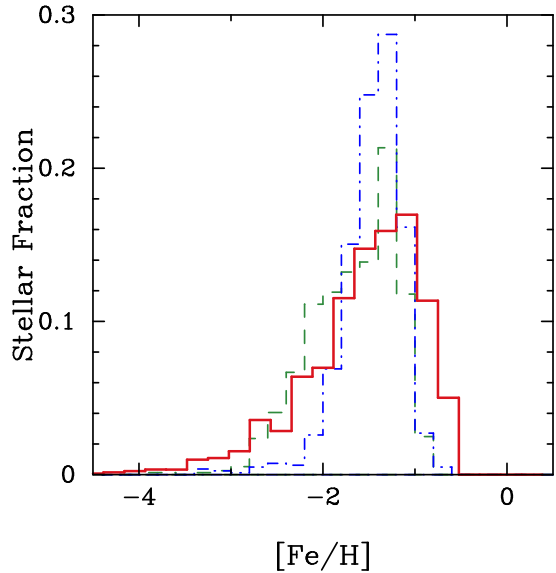}
\caption{The metallicity distribution function of model B (red-solid histogram) at 13.8 Gyr from the beginning of the simulation. The green dashed and blue dotted-dashed histograms denote the observed metallicity distribution functions of the dwarf galaxies Sculptor and Leo I, respectively \citep{2009ApJ...705..328K,2010ApJS..191..352K,2012AJ....144..168K}. \label{MD}}
\end{figure}

Figure \ref{MD} compares the computed and observed MDFs. As shown in this figure, model B has a similar MDF with these galaxies. The {interquartile ranges (IQRs)\footnote{IQR is the difference between the first and third quartiles. The first and third quartiles mean the medians of the lower and upper half of the data set, respectively.} of the MDFs are 0.77 (model B), 0.71 (Sculptor), and 0.36 (Leo I).} Other models have almost the same MDFs because NSMs and ECSNe produce little amount of Fe. Model B exhibits more extended star formation histories compared to those of Sculptor and Leo I, resulting in a more significant fraction of stars for [Fe/H] $>$ $-$1. In this model, the effects of ram-pressure and tidal stripping are not included. These effects may affect the star formation histories in the model. Lack of gas accretion in a cosmological context may cause the more gentle peak of MDF than that of Leo I.  {Mergers of galaxies may also induce additional star formation. However, {this effect is degenerate with} the choice of the initial conditions.} Since this paper aims to discuss the enrichment of Sr mainly at low metallicity, these differences do not affect our conclusions. Although we do not intend to reproduce the chemical evolution of a specific galaxy, we confirm that our models have similar MDFs with those of similar-mass LG dwarf galaxies.

\subsection{Enrichment of Strontium}\label{Sr}
The ratio of [Sr/Fe] is a clear indicator of the enrichment of Sr in the LG galaxies. In the MW, there are star-to-star scatters in [Sr/Fe] for [Fe/H] $<$ $-$2.0 (Figure \ref{SrFeobs}). At higher metallicity, the [Sr/Fe] evolution becomes flat with a mean value of 0.06. In the dwarf galaxy Sculptor ($M_{*}~{\sim}~10^6 M_{\sun}$), there seems to be an increasing trend of [Sr/Fe] ratios toward higher metallicity. One exception is a star, ET0381 ([Fe/H] = $-$2.44, [Sr/Fe] = $-$0.76). The subsolar ratios of $\alpha$-elements as well as low Ni and Cr abundances of this star suggest that it was formed from the gases polluted by few massive stars \citep{2015A&A...583A..67J}. In the dwarf galaxies Draco and Ursa Minor ($M_{*}~{\sim}~10^5 M_{\sun}$), there are relatively flat trends of [Sr/Fe] ratios compared to that of Sculptor. Most of UFDs with $M_{*}~\lesssim~10^5 M_{\sun}$ are depleted in [Sr/Fe] \citep{2015ARA&A..53..631F}. A clear jump in the [Sr/Fe] evolution seen in the UFD Reticulum II is possibly due to the contribution of an NSM \citep{2016Natur.531..610J, 2016ApJ...830...93J, 2016AJ....151...82R, 2018ApJ...865...87O}. Stars enhanced in [Sr/Fe] are also confirmed in Canes Venatici I and II \citep{2016A&A...588A...7F}.

\subsubsection{Contribution of NSMs and AGBs to the Enrichment of Sr}\label{NSM}
\begin{figure*}[htbp]
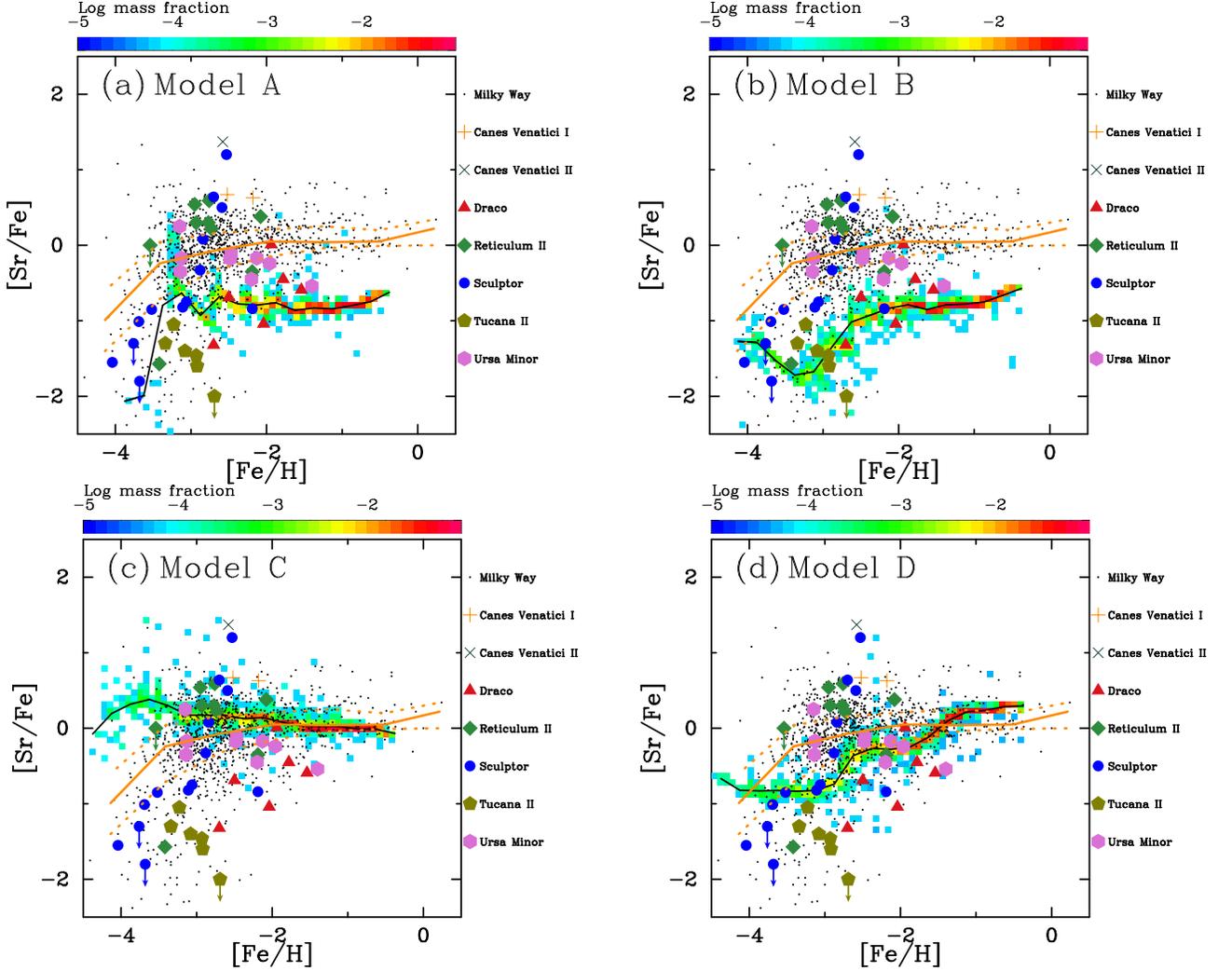

\epsscale{1.0}
\gridline{\fig{f3a_r2.eps}{0.45\textwidth}{}
\fig{f3b_r2.eps}{0.45\textwidth}{}
}
\vspace{-1.0cm}
\gridline{
\fig{f3c_r2.eps}{0.45\textwidth}{}
\fig{f3d_r2.eps}{0.45\textwidth}{}
}
\vspace{-0.8cm}
\caption{Stellar [Sr/Fe] ratios as a function of [Fe/H] at 13.8 Gyr from the beginning of the simulation in models A (panel a), B {(panel b), C (panel c), and D (panel d)}. The color coding from blue to red denotes the mass fractions of stars computed in our models from 10$^{-5}$ to 10$^{-1}$ in each grid of 0.08 $\times$ 0.08 dex$^2$. When the grid contains one star, the resulting mass fraction is 8.8 $\times$ 10$^{-5}$ (sky-blue in the color-coding). The black solid curve shows the median of [Sr/Fe] ratios predicted by the model. The orange solid and dotted curves denote the median and the first and third quartiles of the [Sr/Fe] ratios in the MW. The dots denote the observed abundances in the MW. The symbols represent stellar abundances in selected dwarf galaxies: Canes Venatici I (orange pluses), Canes Venatici II (light sea green crosses), Draco (red triangles), Reticulum II (green diamonds), Sculptor (blue circles), Tucana II (olive green pentagons), and Ursa Minor (magenta hexagons). Data with upper limits are shown with arrows.\label{SrFe}}
\end{figure*}

NSMs contribute to the enrichment of Sr. In model A, NSMs eject Sr for [Fe/H] $\gtrsim$ $-$3. Figure \ref{SrFe}a represents [Sr/Fe] as a function of [Fe/H] in model A. As shown in this figure, Sr from an NSM appears at [Fe/H] $\sim$ $-$3.\footnote{Note that the random seed {affects the onset of NSM events (see Appendix \ref{ap:NSMs}}).}. NSMs give rise to gaps in the average ratios of [Sr/Fe]. In model A, the first NSM occurs at {1.23} Gyr from the beginning of the simulation. The mean [Sr/Fe] ratio in the gas phase jumps from {$-$2.2 to $-$0.7} when the first NSM occurs. {The very low [Sr/Fe] ratios ($\sim -2$) at the lowest metallicity come from AGBs.}

Such increases in the abundances of neutron-capture elements can be seen in LG dwarf galaxies. In the dwarf galaxy Draco, the mean [Ba/H] and [Y/H] ratios increase from $-$2.89 to $-$1.92 and $-$3.5 to $-$2.5, respectively \citep{2017ApJ...850L..12T}. The UFD Reticulum II also shows a similar behavior \citep{2016Natur.531..610J, 2016ApJ...830...93J, 2016AJ....151...82R}. Our result suggests that such gaps in the abundances of neutron-capture elements are the signatures of the first NSMs occurred in dwarf galaxies.

For [Fe/H] $>$ $-$1, the [Sr/Fe] ratio increases owing to the contribution of AGBs. Because of the strong metallicity dependence on the yields of Sr, the contribution of AGBs in the enrichment of Sr is subdominant at low metallicity \citep{2018MNRAS.476.3432P}. Note that AGBs can eject materials from [Fe/H] $\gtrsim$ $-$4 because of the slow chemical evolution in dwarf galaxies. However, owing to the deficit of seed nuclei, the amount of Sr from AGBs is too small to enrich the gas at low metallicity. Despite the contribution of AGBs, the mean value of [Sr/Fe] ratios in model A  is $\sim$ 1 dex lower than that of the MW halo. This discrepancy implies that additional sources such as ECSNe \citep[e.g.,][]{2018ApJ...852...40W, 2019arXiv190802236J} or RMSs \citep[e.g.,][]{2018ApJS..237...13L} contributes to the enrichment of Sr.

\subsubsection{Contribution of ECSNe to the Enrichment of Sr}
Figure \ref{SrFe}{b} shows the [Sr/Fe] ratios as a function of [Fe/H] in model {B}. Stars with Sr for [Fe/H] $<$ $-$3.5 originate from ECSNe, because the first NSM occurs at [Fe/H] = $-2.6$ in this model. The mean value of [Sr/Fe] ratios in model {B} is [Sr/Fe] = $-$1.53 for [Fe/H] $<$ $-$3.5. Stars in Sculptor and UFDs are also depleted in [Sr/Fe] at the low-metallicity end \citep[e.g.,][]{2015ARA&A..53..631F, 2019ApJ...870...83J}. Due to the low rate of ECSNe, their ejecta that are well mixed into surrounding gases without enrichment by ECSNe result in the low ratios of [Sr/Fe]. We find that model {B} forms stars down to [Sr/Fe] $\sim$ $-$4 in this metallicity range. This result suggests that stars at the lowest metallicity were formed from the well-mixed gas containing the ejecta from ECSNe. Note that the scatters of [Sr/Fe] ratios tend to be underestimated owing to isolated evolution of our models. Gas accretion in the cosmological context may induce more scatters.

 The overall lower median of [Sr/Fe] ratios than that in the MW can be attributed to the uncertainties in the mass range of ECSNe (Section \ref{ECSN}) or the rate and yields of NSMs (Section \ref{NSM}), as well as the contribution of an additional nucleosynthetic site (Section \ref{MS}). Since we adopt isolated dwarf galaxy models, the scatters of [Sr/Fe] ratios in our models also are smaller than those of the MW. As suggested by \citet{2018ApJ...865...87O}, the scatters in the abundances of neutron-capture elements can be explained when the MW halo is formed from the clustering of different mass galaxies. If this is the case, the smaller scatters of [Sr/Fe] ratios in model {B} than those in the MW halo are reasonable in our isolated dwarf galaxy model.

\begin{figure}[htbp]
\epsscale{1.3}
\plotone{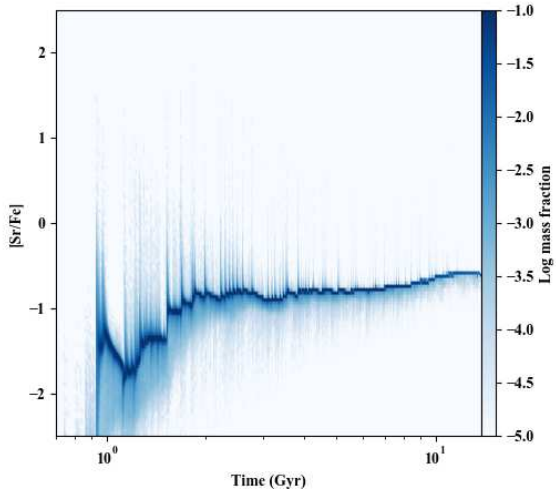}
\caption{Gas-phase [Sr/Fe] ratios in model {B} within one kpc from the center of the galaxy as a function of time, in which most of the stars are formed. The mass fraction of gases in each bin is displayed with the color scale from white (10$^{-5}$) to blue (10$^{-1}$). The bin size corresponds to the length of axes divided by 128. \label{SrFegas}}
\end{figure}

Stars appreciably affected by the ejecta from ECSNe or NSMs have enhanced [Sr/Fe] ratios. Figure \ref{SrFegas} shows the time evolution of the gas phase [Sr/Fe] ratios in model {B}. Discrete events of ECSNe and NSMs produce several strong peaks in the [Sr/Fe] ratios. Metal mixing moderates the high [Sr/Fe] ratios in the gas phase. \citet{2017ApJ...838L..23H} estimated the timescale of metal mixing to be $\lesssim$ 40 Myr, which is comparable to the dynamical time of molecular clouds \citep[$n_{\rm{H}}$ $\sim$ 1000 cm$^{-3}$,][]{1981MNRAS.194..809L}. This result suggests that stars enhanced in [Sr/Fe] are formed before the metal mixing moderates the gases with high [Sr/Fe] ratios.

\subsubsection{Dependence on the Progenitor Mass Range of ECSNe}\label{ECSN}
The mass range of the progenitors of ECSNe is highly uncertain, which is affected by the treatment of the mass-loss, third dredge-up, and overshooting \citep[e.g.,][]{2017PASA...34...56D, 2018PASA...35...38G}. \citet{2007PhDT.......212P} predicted a wider mass range of ECSNe than that in \citet{2015MNRAS.446.2599D} using different parameters in stellar evolution models. Stars with striped envelopes in close binary systems can also become ECSNe \citep{2013ApJ...778L..23T, 2015MNRAS.451.2123T, 2016MNRAS.461.2155M}. Binary evolution calculations predict that the resultant mass range of ECSNe can be wider than that for single systems \citep{2004ApJ...612.1044P, 2017ApJ...850..197P}. Besides, \citet{2018ApJ...852...40W} have shown that the iron-core CCSNe in the lowest mass range can synthesize light trans-iron elements with almost the same amounts to those from ECSNe. These uncertainties influence the enrichment of Sr in our model.

Figure \ref{SrFe}{c} shows the [Sr/Fe] ratios as functions of [Fe/H] for model {C}. This model assumes that stars with 8.2--9.2 $M_{\sun}$ at the metallicity greater than 10$^{-5}$ $Z_{\sun}$ explode as ECSNe. According to these figures, the median [Sr/Fe] ratios are larger than that of model {B}. Model {C} appears consistent with the stellar values in the MW and some dwarf galaxies for [Fe/H] $> -3$ but sizably overproduces Sr at lower metallicity. These results suggest that the mass ranges of ECSN progenitors should be appreciably greater than that in model {B} \citep[predicted by][]{2015MNRAS.446.2599D} at higher metallicity to account for the mean [Sr/Fe] ratios in the LG dwarf galaxies or the MW. Although such a possibility may not be excluded, it opposes to the theoretical predictions that the mass range of ECSNe tends to be smaller at higher metallicity owing to more effective mass loss.

\subsubsection{Contribution of RMSs to the Enrichment of Sr}\label{MS}
Previous nucleosynthetic studies have proposed some possible mechanisms to synthesize Sr in massive stars. The innermost ejecta of CCSNe have long been thought as the production site of heavy elements including Sr. Recent studies based on sophisticated hydrodynamical simulations suggest, however, that the bulk of innermost ejecta become proton-rich, in which few heavy elements are produced \citep[except for ECSNe and lowest-mass CCSNe,][]{2018ApJ...852...40W}. Another possible mechanism is the weak $s$-process in RMSs \citep[e.g.,][]{2018MNRAS.476.3432P, 2018ApJS..237...13L}. Elements beyond Zn are synthesized owing to the neutron production by the reaction $^{22}$Ne($\alpha$, $n$)$^{25}$Mg. In RMSs, stellar rotation mixes matter between the He convective core and the H burning shell. This mixing brings carbon made by the 3$\alpha$ reaction to the base of the H shell, resulting in the production of $^{14}$N. This $^{14}$N is brought to the core and converted to $^{22}$Ne, which is the source of free neutrons.

Contribution of RMSs can affect both the median and scatters of [Sr/Fe] ratios. Figure \ref{SrFe}{d} shows the [Sr/Fe] ratios as a function of [Fe/H] in model D, assuming that the RMSs with the rotation speed of 150 kms$^{-1}$ contribute to the enrichment of Sr. As shown in this figure, the median [Sr/Fe] ratios is $\sim1$ dex higher than that of model B (Figure \ref{SrFe}{b}). This result implies that the contribution of Sr from RMSs may account for the deficiency of [Sr/Fe] in model B. Note that the evolutionary history of [Sr/Fe] depends on the rotational velocity distribution (that we do not consider in this study), {although an increasing trend of [Sr/Fe] ratios with metallicity} appears robust \citep{2018MNRAS.476.3432P}.

\begin{figure}[htbp]
\epsscale{1.3}
\plotone{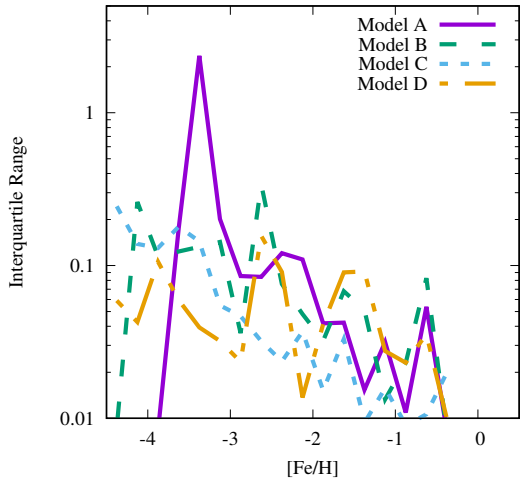}
\caption{Interquartile range of [Sr/Fe] ratios as a function of [Fe/H] in models A (purple solid curve) , B {(green dashed curve), C  (sky-blue dotted curve), and D (orange dot-dashed curve)}. \label{IQR}}
\end{figure}

Figure \ref{IQR} compares the IQRs of [Sr/Fe] as a function of [Fe/H] in models A, B, {C,} and D. In model A, {the first NSM makes large scatters (IQR $>$ 1.0) at [Fe/H] $<$ $-$3.} The contribution of ECSNe also makes IQR up to 0.35, as shown in model B (green-dashed curve). Model C produces {scatters similar or smaller than those of model B}. On the other hand, model D shows almost no scatters at [Fe/H] $<$ $-$3. The first NSM that occurs at [Fe/H] = $-$2.8 temporary increases the IQR but the overall values of IQR are less than 0.1. In model D, all CCSNe (resulting from RMSs) produce Sr, while ECSNe produce Sr with a rate of $\approx$3\% of all CCSNe in model {B}. Frequent production of Sr from massive stars erases scatters of Sr. This result means that the [Sr/Fe] ratios in dwarf galaxies should exhibit small scatters in [Sr/Fe] if RMSs predominantly contribute to the enrichment of Sr. Note that this model tends to suppress the scatters of [Sr/Fe], because we adopt the SSP approximation for each star particle.

AGBs may also produce scatters because their ejecta are possibly less efficiently mixed compared to the ejecta from SNe \citep{2018ApJ...869...94E}. However, it would be {hard} to identify the scatters caused by AGBs in our models. This is because the {spatial} metallicity distribution has been already homogenized after AGBs contribute to the enrichment of Sr.

We find that both ECSNe and RMSs can be the dominant sources of Sr, which tend to be more efficient contributors at lower and higher metallicities, respectively. The values of [Sr/Fe] and its scatters (or IQRs) can be the indicators to distinguish the contributions of these sites. In model B, the mean [Sr/Fe] ratio and IQR at [Fe/H] $=$ $-$4 are $-$1.35 and 0.26, respectively. On the other hand, model D shows higher [Sr/Fe] ratios ($\approx$ $-$0.85) and lower IQR (= 0.05). Higher [Sr/Fe] ratios are owing to the addition of RMSs to model B. Note that these values highly depend on the mass range of ECSNe and the rotational velocity of RMSs. 

It is difficult to distinguish, however, the contributions of these sources from the observed data presently available. The Sr abundances are measured for less than ten stars in each LG dwarf galaxy. The typical observational error is also as large as 0.2 dex, which is comparable to the degree of scatters caused by ECSNe. It is necessary to perform observations with approximately 100 stars at [Fe/H] $<~-$2 with observational errors less than 0.1 dex in dwarf galaxies to distinguish these astrophysical sites of Sr.

\subsubsection{Enrichment of Sr without the Contribution of NSMs}\label{sec:noNSM}
\begin{figure*}[htbp]
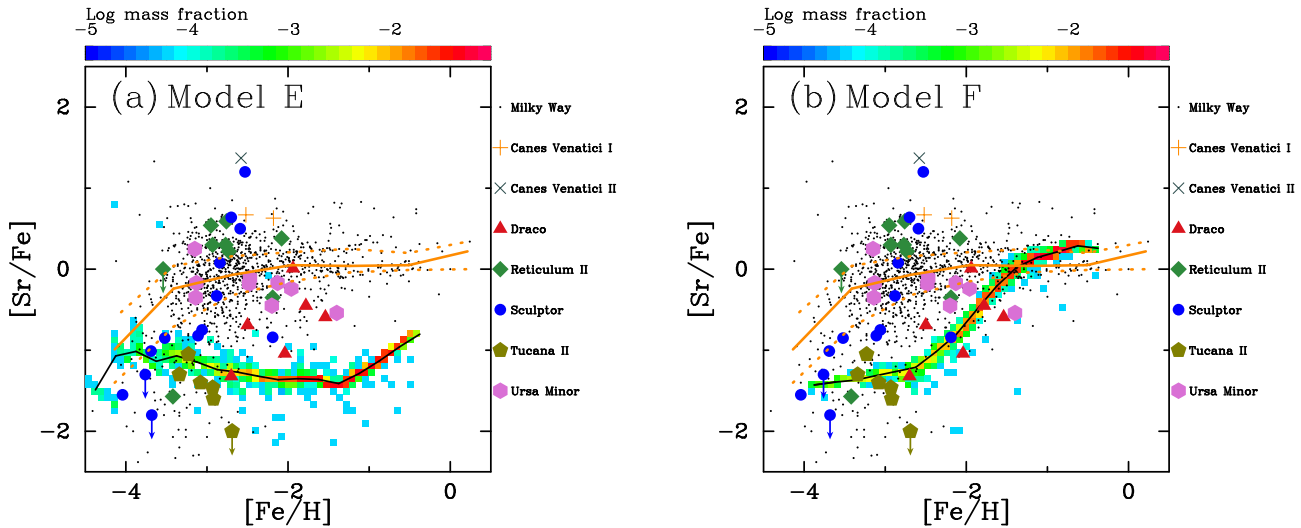

\epsscale{1.0}
\gridline{\fig{fB1a.eps}{0.45\textwidth}{}
\fig{fB1b.eps}{0.45\textwidth}{}
}
\vspace{-0.8 cm}
\caption{Same as Figure \ref{SrFe}, but for [Sr/Fe] as a function of [Fe/H] in models E (panel a) and F (panel b).\label{SrFenoNSM}}
\end{figure*}

{Here we present the simulations excluding NSMs to inspect the contributions of ECSNe and RMSs. Figure \ref{SrFenoNSM} shows the [Sr/Fe] ratios as a function of [Fe/H] in models E and F. These models remove the contribution of NSMs from models B and D, respectively. As shown in this figure, [Sr/Fe] ratios at [Fe/H] $\sim$ $-$1 in model E are about 1 dex lower than those in model B. Both models show similar ratios of [Sr/Fe] (= $-$1.3 in model E and = $-$1.2 in model F) at [Fe/H] = $-$2.5.}

{The major difference between these models is the scatters of [Sr/Fe] ratios. The IQR in model F (=0.02) is substantially smaller than that in model E (=0.6). The trends of [Sr/Fe] evolution are also different. Model E shows flat [Sr/Fe] ratios, while model F has an increasing trend of [Sr/Fe] ratios owing to the metallicity dependence of the yields. This steep increase of [Sr/Fe] can be, however, moderated by the metallicity dependence of the RMS yields on the rotational velocity distribution. Rotational velocity tends to be lower at a higher metallicity environment \citep{2008A&A...479..541H, 2009A&A...496..841H}. The yield of Sr at a given metallicity is smaller in the model with a lower rotational velocity \citep{2018ApJS..237...13L}.}

\section{Discussion}\label{sec:dis}
\subsection{Contribution of the $s$-process}\label{s-process}
We discuss a role of the $s$-process from AGBs and RMSs on the enrichment of Sr. The ratios of [Ba/Eu] can be an indicator of the contribution of the $s$-process. In the solar system abundances, 85\% of Ba originates from the $s$-process and 97\% of Eu from the $r$-process \citep{1989RPPh...52..945K, 2000ApJ...544..302B, 2004ApJ...617.1091S}. The main $s$-process that produces Ba occurs in AGBs \citep{2011RvMP...83..157K}. The weak $s$-process in RMSs also produces a non-negligible amount of Ba \citep{2013ApJ...764...21C, 2018ApJS..237...13L}. Due to its secondary nature as well as the long lifetimes of low and intermediate-mass stars, the $s$-process starts contributing to galactic chemical evolution later than the $r$-process does.

\begin{figure*}[htbp]
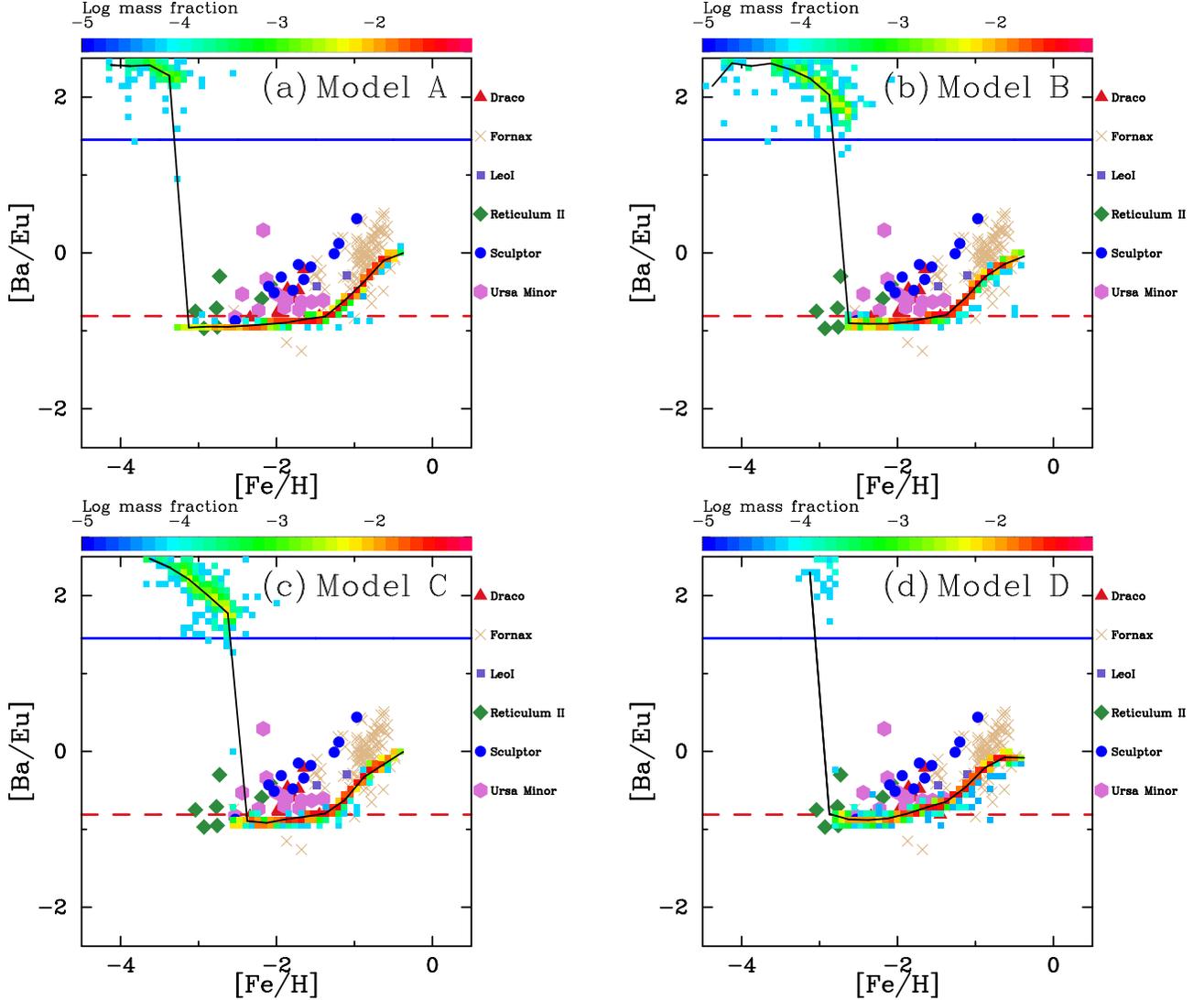

\epsscale{1.0}
\gridline{\fig{f6a_r2.eps}{0.45\textwidth}{}
\fig{f6b_r2.eps}{0.45\textwidth}{}
}
\vspace{-1.0cm}
\gridline{\fig{f6c_r2.eps}{0.45\textwidth}{}
\fig{f6d_r2.eps}{0.45\textwidth}{}
}
\vspace{-0.8cm}
\caption{Same as Figure \ref{SrFe}, but for stellar [Ba/Eu] as a function of [Fe/H] in models A (panel a), B (panel b), {C (panel c), and D (panel d)}. Red dashed and blue-solid lines represent the solar $r$- and $s$-process ratios, respectively \citep{2004ApJ...617.1091S}. {Purple squares represent the observed data of Leo I \citep{2003AJ....125..684S}. The references and descriptions of symbols for the other galaxies are given in the captions of Figures \ref{SrFeobs} and \ref{SrFe}, respectively.}\label{BaEu}}
\end{figure*}

Figure \ref{BaEu} shows the stellar [Ba/Eu] ratios as a function of [Fe/H] in models A, B, {C} and D. According to this figure, the input parameters of ECSNe and RMSs do not greatly affect the [Ba/Eu] ratios. Stars with [Ba/Eu] $>$ 1 for [Fe/H] $\lesssim-$2 reflect the ejecta of AGBs. The mass loss timescale of AGBs assumed in this model is $\approx$ 0.3 Gyr. Because of the slow chemical evolution in this model, AGBs can eject materials even at [Fe/H] $\lesssim-$3. However, these stars with high [Ba/Eu] ratios have too low Ba and Eu abundances ([Ba/H] $<-$6 and [Eu/H] $<-$4) to be detected in the current observations. After the first NSM occurs, most of the stars reside near the solar $r$-process ratios ($-$3$<$ [Fe/H] $<-$2). On the other hand, the [Ba/Eu] ratios start increasing at [Fe/H] $\sim$ $-$1.5 in our models. This result means that AGBs have an appreciable contribution to the enrichment of Ba for [Fe/H] $>$ $-$1.5  (and also of Sr for [Fe/H] $>$ $-$1; see Figure \ref{SrFe}). Note that the [Ba/Eu] ratio in model D starts to increase at slightly lower metallicity than those in the other models owing to the non-negligible production of Ba in RMSs.

As recently reported by \citet{2018arXiv181201486H, 2019arXiv190810729S} using the VLT/FLAMES high-resolution spectroscopy, there is a clear increase of [Ba/Eu] ratios from [Fe/H] {$\approx -2.0$} in Sculptor. \citet{2018ApJ...869...50D} obtained the slopes of [Ba/Eu] as functions of [Fe/H] be 0.59 in Sculptor and 0.58 in Fornax, indicating that the $s$-process increases [Ba/Eu] ratios at higher metallicity. On the other hand, stars in Ursa Minor still resides near the pure $r$-process ratio at [Fe/H] = $-$1.5. The metallicity at which the [Ba/Eu] ratio starts to increase depends on the timescale of chemical evolution. These results imply that our models increase iron abundances with similar timescale in Ursa Minor. Sculptor possibly evolved slower than our models.

\subsection{Ratios of Light to Heavy Neutron-Capture Elements}\label{sites}
The [Sr/Ba] ratios provide us with some hints for the astrophysical sites of light and heavy neutron-capture elements. Since ECSNe and RMSs mainly synthesize light trans-iron elements (including Sr) with no and a small amount of Ba, respectively, the stars that inherit the abundances of their ejecta have enhanced [Sr/Ba] ratios. On the other hand, the stars that directly inherit the abundances of ejecta from NSMs have lower [Sr/Ba] ratios.

\begin{figure*}[htbp]
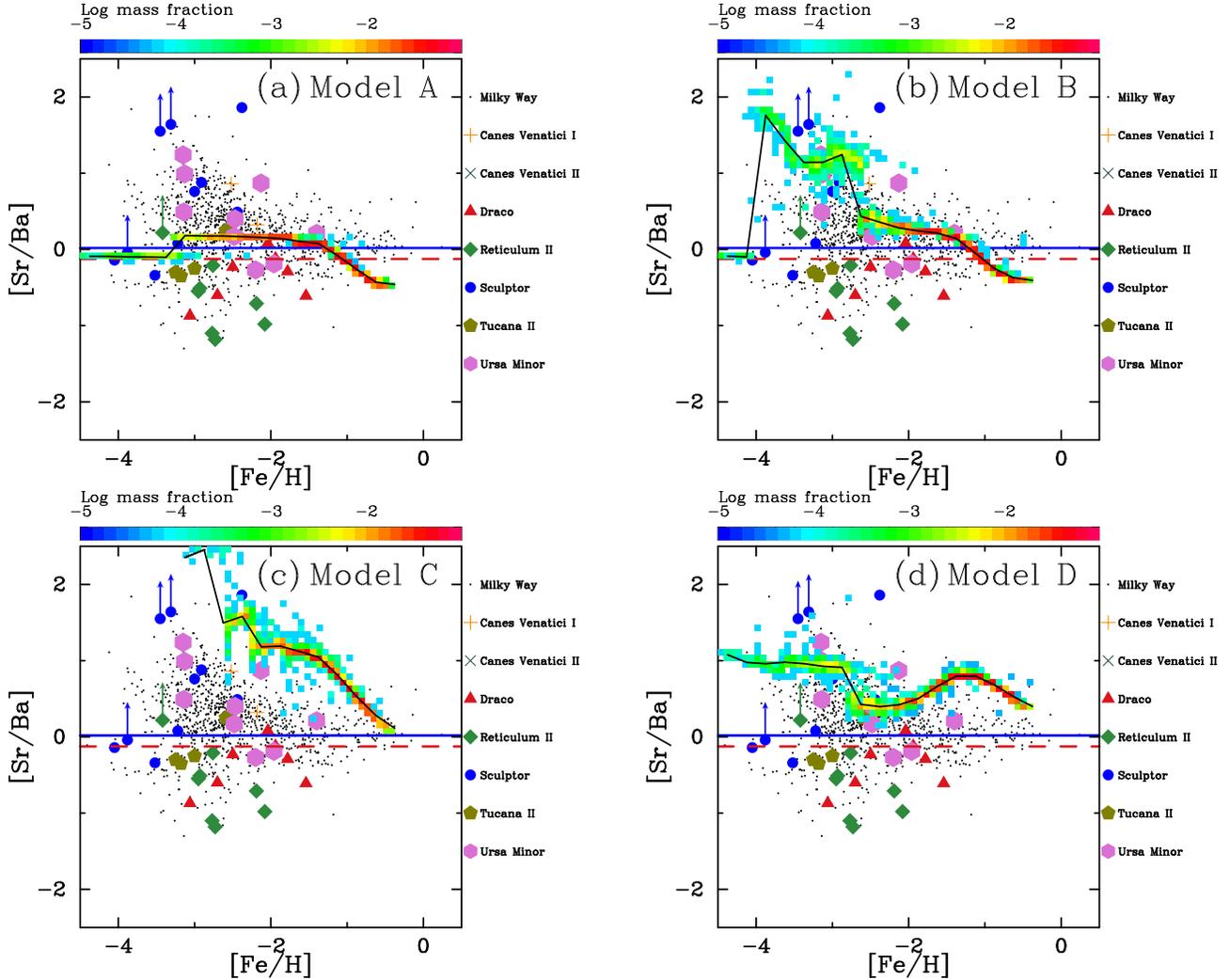

\epsscale{1.0}
\gridline{\fig{f7a_r2.eps}{0.45\textwidth}{}
\fig{f7b_r2.eps}{0.45\textwidth}{}
}
\vspace{-1.0cm}
\gridline{\fig{f7c_r2.eps}{0.45\textwidth}{}
\fig{f7d_r2.eps}{0.45\textwidth}{}
}
\vspace{-0.8cm}
\caption{Same as Figure \ref{SrFe}, but for [Sr/Ba] as a function of [Fe/H] in models A {(panel a), B (panel b), {C (panel c), and D (panel d)}}. Stars that only have the upper limit of Ba abundance are shown by arrows. Red-dashed and blue-solid lines represent the solar $r$- and $s$-process ratios, respectively \citep{2004ApJ...617.1091S}. \label{SrBa}}
\end{figure*}

Figure \ref{SrBa} shows the [Sr/Ba] ratios as a function of [Fe/H] in models A, B, {C}, and D. In our models, the [Sr/Ba] enhanced stars directly reflect the ejecta from ECSNe (models {B} and C) and RMSs (model D). In models B, {C}, and D, 5.2, 48.1 and 5.9 \% of all stars have [Sr/Ba] $\geq$ 1. On the other hand, we find that 4.5 \% of all stars except for carbon-enhanced stars have [Sr/Ba] $\geq$ 1.0 in the MW \citep[SAGA database,][]{2008PASJ...60.1159S, 2011MNRAS.412..843S, 2013MNRAS.436.1362Y}. These results indicate that model {C} with the large mass window of ECSN progenitors ($\Delta M = 1\, M_\odot$) is not compatible with the MW or LG dwarf galaxies. The over-prediction of [Sr/Ba] at [Fe/H] $> -2$ in model D implies a smaller contribution of RMSs (i.e., a smaller rotational velocity) at higher metallicity in reality as suggested from theoretical \citep{2013ApJ...764...21C, 2018MNRAS.476.3432P, 2018ApJS..237...13L} and observational \citep{2008A&A...479..541H, 2009A&A...496..841H} studies.

There are several stars with highly enhanced [Sr/Ba] ratios in the LG dwarf galaxies. The most notable one is SMSSJ022423.27-573705.1 reported by \citet{2015ApJ...807..171J}. This star has a large [Sr/Fe] ratio ([Sr/Fe] = 1.08) but no detectable Ba line ([Ba/Fe] $<$ $-$0.91), indicating that the [Sr/Ba] ratio is greater than 2.0. There are also clear signatures of the additional sources of Sr in the LG dwarf galaxies. As reported by \citet{2016A&A...588A...7F}, the dwarf galaxy Canes Venatici II has a star with [Sr/Ba] $>$ 2.6. Although they can only put the upper limit of [Ba/Fe], the ratio of [Sr/Fe] of this star is 1.37. They also show that the dwarf galaxy Canes Venatici I has stars with [Sr/Ba] = 0.22 and 0.76. Our results suggest that the presence of such [Sr/Ba] enhanced stars can be explained if they are formed from the gas polluted by ECSNe.

{It should be noted that, in} our models (Figure \ref{SrBa}), we adopt the {single} yield table of \citet{2014ApJ...789L..39W} for NSMs, which results in {a constant value of} [Sr/Ba] = 0.187. Therefore, stars below {or above} [Sr/Ba] = 0.187 cannot be formed from the ejecta of NSMs {(Figure \ref{SrBa}a)}. However, the nucleosynthesis of light $r$-process elements in NSMs is still uncertain. These elements are formed in the relatively less neutron-rich ejecta from NSMs \citep{2014ApJ...789L..39W}. If stars were formed reflecting the nucleosynthetic yields from less neutron-rich ejecta \citep[that is the case for NSMs with smaller mass ratios,][]{2016PhRvD..93l4046S}, the [Sr/Ba] ratios would become higher. Besides, the post-merger outflows from the accretion disks in binary neutron stars can synthesize $r$-process elements \citep[e.g.,][]{2015MNRAS.448..541J, 2016MNRAS.463.2323W, 2017PhRvL.119w1102S, 2018ApJ...860...64F}. The neutron richness of the post-merger ejecta depends on the models (e.g., if the central remnant is a massive neutron star or a black hole).

The lower ratios of [Sr/Ba] in several stars than the solar $r$-process ratio can also be due to the contribution from binary black hole-neutron star (BH-NS) mergers \citep{1974ApJ...192L.145L, 1976ApJ...210..549L}. Since tidal torques are the main driver of the mass ejection, BH-NS mergers may eject very neutron-rich matter, which results in the production of heavy (but little amounts of light) $r$-process elements \citep[e.g.,][]{2013ApJ...776...47D, 2017CQGra..34o4001F, 2018PhRvD..97b3009K}. 

Spectroscopic studies of MW halo stars indicate, however, only a small variation of Sr/Ba ratios (within a factor of several) among those with high $r$-process enhancement \citep[{more than 10 stars,}][]{2019arXiv190101410C}. Given these stars reflecting single $r$-process events, the stars with [Sr/Ba] $> 1$ (or $< 1$) {may not} be explained by the astrophysical source of the $r$-process (either NSMs or BH-NS mergers). {Note that the number of observed {$r$-process-enhanced} stars is still too small to exclude a possible contribution of NSMs to such high (or low) stellar [Sr/Ba] ratios.}

\subsection{Ratios of Sr to Zn}\label{sec:SrZn}
The astrophysical site of the heaviest iron-group element, Zn, is not well understood, either. \citet{2018ApJ...855...63H} have shown that ECSNe can contribute to the enrichment of Zn. HNe are also a possible site of Zn \citep{2002ApJ...565..385U, 2005ApJ...619..427U, 2006ApJ...653.1145K, 2007ApJ...660..516T}. Besides these possibilities, \citet{2018ApJ...863L..27T} suggested that magnetorotational SNe could contribute to the production of Zn at low metallicity.

Spectroscopic studies of the MW halo stars have shown that there is a decreasing trend of [Zn/Fe] toward higher metallicity with small scatters for [Fe/H] $\lesssim$ $-$2 \citep[e.g.,][]{2004A&A...416.1117C, 2004A&A...415..993N, 2007A&A...469..319N, 2009PASJ...61..549S,2017A&A...604A.128D}. The LG dwarf galaxies appear to have similar behaviors \citep[e.g.,][]{2010ApJ...708..560F, 2010ApJ...719..931C,2012ApJ...751..102V, 2017A&A...606A..71S, 2018A&A...615A.137S}. The {IQR of [Zn/Fe] ratios for [Fe/H] $<$ $-$2 in the MW halo is $\approx$ 0.2} \citep[SAGA database,][]{2008PASJ...60.1159S, 2011MNRAS.412..843S, 2013MNRAS.436.1362Y}. This means that the scatters are within the observational errors. In contrast, [Sr/Fe] ratios have large scatters (IQR = 0.5) in the MW halo as described in Section \ref{Sr}. These different levels of scatters may help us understand the astrophysical sites of these elements.

\begin{figure*}[htbp]
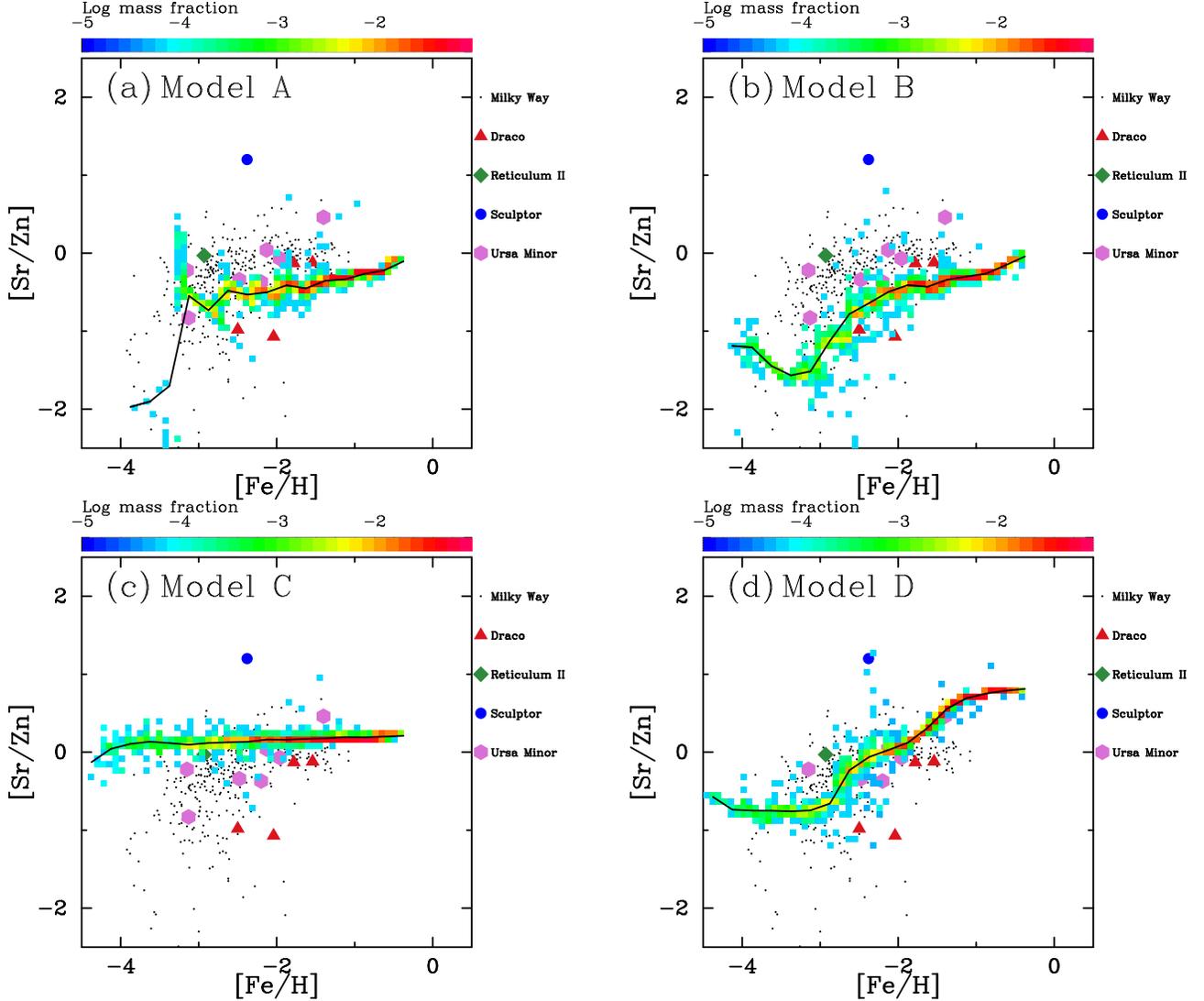

\epsscale{1.0}
\gridline{\fig{f8a_r2.eps}{0.45\textwidth}{}
\fig{f8b_r2.eps}{0.45\textwidth}{}
}
\vspace{-1.0cm}
\gridline{\fig{f8c_r2.eps}{0.45\textwidth}{}
\fig{f8d_r2.eps}{0.45\textwidth}{}
}
\vspace{-0.8cm}
\caption{Same as Figure \ref{SrFe}, but for [Sr/Zn] as a function of [Fe/H] in models A (panel a), B (panel b), {C (panel c), and D (panel d)}.\label{SrZn}}
\end{figure*}

Figure \ref{SrZn} shows the [Sr/Zn] ratios as a function of [Fe/H] in models A, B, {C}, and D. As shown in this figure, the [Sr/Zn] ratios in models {B} and D are overlapped with those in observations. We find that the delayed production of Sr from NSMs makes scatters in these models. NSMs and RMSs make the increasing trend of [Sr/Fe] toward higher metallicity in models {B} and D, respectively. The $s$-process contribution from AGBs further increases the [Sr/Zn] ratios at high metallicity (see also Section \ref{s-process}). The {IQRs of [Sr/Zn] for [Fe/H] $<$ $-$2 are 0.4 and 0.1 in models B and D. On the other hand, the IQR in the observations of the MW halo is 0.6.} As discussed in Section \ref{Sr}, the smaller scatter in our model of an isolated system may be reasonable. The [Sr/Zn] ratio in model {C} (Figure \ref{SrZn}{c}) is almost constant. This is because the large contribution of both Sr and Zn from ECSNe dominates over those from the other sources in this model. As such behavior is incompatible with the increasing trend of [Sr/Zn] in observations, the mass window of ECSN progenitors should be substantially narrower than that ($\Delta M = 1\, M_\odot$) assumed in model {C}.

\subsection{Enrichment of Sr in LG dwarf galaxies}\label{Sculptor}
The chemical evolution of Sr behaves differently among dwarf galaxies with different masses. Stars in several UFDs such as Canes Venatici I, II, Hercules, and Triangulum II are enhanced in [Sr/Ba] \citep{2016A&A...588A...7F, 2017ApJ...838...83K}. As shown in Section \ref{sites}, 5.2\% and 5.9\% of stars have [Sr/Ba] $>$ 1 in models {B} and D, respectively, reflecting the ejecta from ECSNe with Sr but {no} Ba (note that RMSs give [Sr/Ba] $\approx 1$ at low metallicity). These results suggest that stars enhanced in [Sr/Ba] found in UFDs are also due to the ejecta from ECSNe.  If ECSNe occur with a rate of a few percents of all CCSNe, the expected number of ECSNe is larger than unity in UFDs. This result is also consistent with the statistical analysis in \citet{2013ApJ...774..103L}, which suggests the strong mass dependence of the yield of Sr. ECSNe reside only in the lowest mass range of CCSN progenitors.

Most of the observed UFDs are depleted in [Sr/Fe] \citep{2015ARA&A..53..631F}. According to Figure \ref{SrBa}, the [Sr/Ba] ratios in Tucana II are located close to the pure $r$-process ratio. Bo\"otes I also shows a similar trend \citep{2014A&A...562A.146I}. These stars are unlikely to have been formed from the gas enriched by ECSNe or RMSs with high [Sr/Ba]. The rate of NSMs is too low to occur in a majority of UFDs. These stars may have formed from the gases externally enriched by the neighboring halos \citep{2016ApJ...830...76K, 2017ApJ...848...85J}.

The trends of [Sr/Fe], [Sr/Ba], and [Sr/Zn] evolution seen in Draco and Ursa Minor ($M_{*}~{\sim}~10^5 M_{\sun}$) are similar to our models. Models {B} and D reproduce the flat trend of [Sr/Fe] ratios for [Fe/H] $\gtrsim$ $-$2 in these galaxies. These models also show a jump of [Sr/Fe] ratios at low metallicity as seen in Draco, owing to the delay of an NSM contribution. The [Sr/Ba] ratio in Draco increases in the range [Fe/H] $<$ $-$2 but decreases at higher metallicity \citep{2009ApJ...701.1053C}. Our results suggest that ECSNe and RMSs increase the ratios of [Sr/Ba] for [Fe/H] $<$ $-$2 in Draco. At higher metallicity, the contribution of NSMs decreases the [Sr/Ba] ratios.

The trends of [Sr/Fe] and [Sr/Ba] ratios in Sculptor are difficult to reproduce in our models, but these trends may show the signatures of ECSNe. At the lowest metallicity, stars reside near the pure $r$-process ratio of [Sr/Ba] (Figure \ref{SrBa}). Such stars may have formed in similar environments with the low ratios of [Sr/Fe] in UFDs discussed above. The very high stellar values of [Sr/Ba] ($> 1$) need a non-negligible contribution of ECSNe, which cannot be explained by the other sources {(with adopted yields)} considered in this study. As discussed in Section \ref{sec:SrZn}, the stellar abundances of [Sr/Zn] serve to constrain the relative contribution of ECSNe. For Sculptor, there is only one star (except for carbon stars) reported with measurements of both Zn and Sr \citep{2012AJ....144..168K}. It is essential to increase the number of stars with both measured Sr and Zn abundances in the LG dwarf galaxies to confirm this scenario.

\section{Conclusions}\label{sec:con}
In this study, we performed a series of $N$-body/SPH simulations of dwarf galaxies with the contributions from NSMs, AGBs, ECSNe, and RMSs as the sources of Sr. We have shown that ECSNe (or iron-core CCSNe at the low-mass end), NSMs, and RMSs significantly contribute to the enrichment of Sr over a wide range of metallicity, while AGBs play a role only at high metallicity. We also discussed the relations among the abundances of Sr, Ba, and Zn to disentangle relative contributions from the different sources to the enrichment of Sr.

We have found that ECSNe can contribute to the enrichment of Sr mainly in the low metallicity range, [Fe/H] $\lesssim$ $-$3. Due to the lower event rates compared to normal CCSNe (including RMSs), ECSNe give rise to the scatters (IQR $\approx$ 0.2) of [Sr/Fe] ratios. The [Sr/Fe] ratios are significantly affected by the uncertain mass range of ECSN progenitors. We have shown that, even for the models ({B} and D) with a narrow mass range as suggested with recent studies \citep[$\Delta M \sim 0.1$--$0.2\, M_\odot$,][]{2015MNRAS.446.2599D}, the ejecta from ECSNe can lead to the formation of stars with Sr for [Fe/H] $\lesssim$ $-$3. 
 
In our models, NSMs contribute to the enrichment of Sr for [Fe/H] $\gtrsim$ $-$3. A large amount of Sr from an NSM leads to an abrupt increase in the ratios of [Sr/Fe] at low metallicity. These features are consistent with the observations of Draco and Reticulum II. However, models without RMSs (except for model {C}) under-predict the [Sr/Fe] ratios by $\sim 1$~dex over a wide range of metallicity. It appears that the weak $s$-process in RMSs play a vital role in the enrichment of Sr (model D) as suggested by \citet{2018MNRAS.476.3432P}.
We find that the contribution of the main $s$-process from AGBs becomes dominant only for [Fe/H] $\gtrsim$ $-$1.5, as can be seen in the enrichment history of [Ba/Eu].

Enrichment from rare astrophysical events, ECSNe and NSMs (but {not} RMSs or AGBs), can make scatters in [Sr/Fe], [Sr/Ba], and [Sr/Zn] ratios as observed in LG dwarf galaxies and the MW. We also have found that stars enhanced in [Sr/Ba] ($> 1$) are formed from the gas enriched by ECSNe. These results suggest that, despite an appreciable contribution of RMSs across a wide metallicity range, ECSNe and NSMs are necessary to consistently explain the presence of stars with large abundance scatters as well as with enhanced [Sr/Ba] ratios. The mass window for the ECSN channel should, however, be $\Delta M \ll 1\, M_\odot$ to avoid over-prediction of [Sr/Ba] and [Sr/Zn] ratios at higher and lower metallicities, respectively (model {C}).

In summary, ECSNe (or iron-core CCSNe at the low-mass end), NSMs, RMSs, and AGBs are, at least in part, the astrophysical sources that reasonably account for the mean trends and scatters in [Sr/Fe], [Sr/Ba], and [Sr/Zn] ratios observed in LG dwarf galaxies. It should be noted that the small number of stars with measured Sr in dwarf galaxies hampers quantitative comparison of our models with observations. More observational data of Sr, along with other key elements (such as Zn and Ba) are required to disentangle the relative contributions of these sources. It is also necessary to conduct more detailed nucleosynthesis studies, stellar evolution calculations, and high-resolution cosmological zoom-in simulations to resolve these issues entirely.

\acknowledgements
We are grateful for an anonymous referee who gave us insightful comments. We also thank the participants of the workshop, ``Nucleosynthesis and electromagnetic counterparts of neutron-star mergers" at Yukawa Institute for Theoretical Physics, Kyoto University (No. YITP-T-18-06) and the Lorentz Center workshop ``Electron-capture initiated stellar collapse" for many useful discussions. This work was supported by JSPS KAKENHI Grant Numbers 19K21057, 18H05876, 19H01933, 26400237, 26707007, MEXT SPIRE and JICFuS. YH is supported by the Special Postdoctoral Researchers (SPDR) program at RIKEN. Numerical computations and analysis were carried out on Cray XC30, XC50, and computers at Center for Computational Astrophysics, National Astronomical Observatory of Japan and the 
Yukawa Institute Computer Facility. This research has made use of NASA's Astrophysics Data System.

\appendix
\section{Variations of [Sr/Fe] evolution caused by the stochastic onsets of NSMs}\label{ap:NSMs}
\begin{figure*}[htbp]
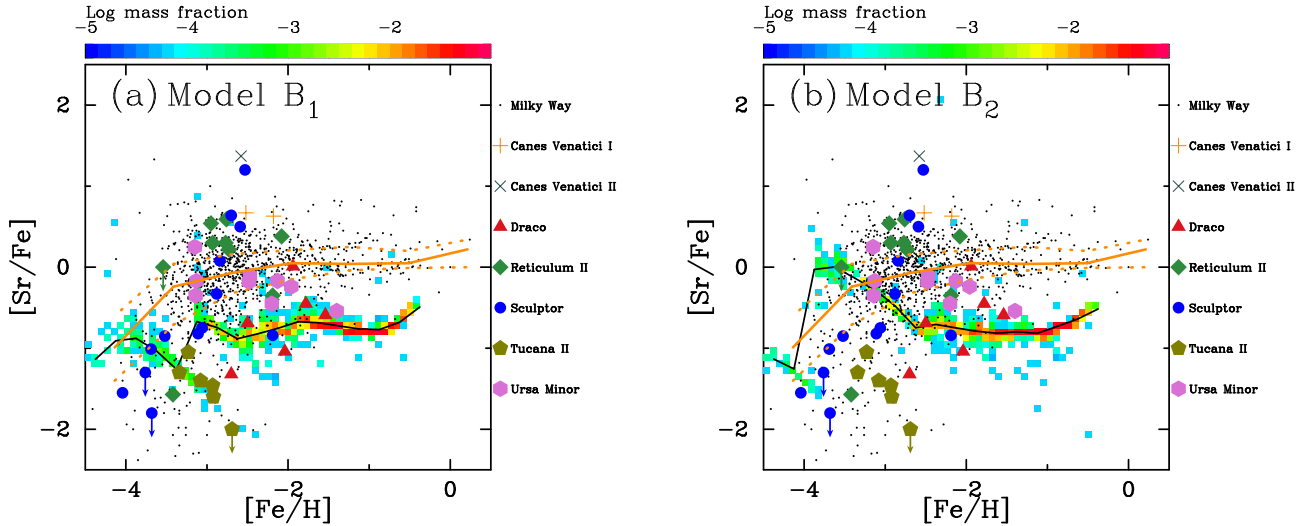

\epsscale{1.0}
\gridline{\fig{fA1a.eps}{0.45\textwidth}{}
\fig{fA1b.eps}{0.45\textwidth}{}
}
\vspace{-0.8 cm}
\caption{Same as Figure \ref{SrFe}, but for [Sr/Fe] as a function of [Fe/H] in models B$_1$ (panel a) and B$_2$ (panel b).\label{SrFeNSM}}
\end{figure*}

The onsets of the NSM contribution at different metallicities make variations of [Sr/Fe] evolution. Figures \ref{SrFeNSM}a and b compare models with the same parameters adopted in model B but with different random number seeds. The first NSMs occur at [Fe/H] = $-$3.0 and $-$4.0 in models B$_1$, and B$_2$, respectively. 

Variations of [Sr/Fe] evolution are also seen in LG dwarf galaxies. Draco shows an increasing trend of [Sr/Fe] ratios, while Ursa Minor has a slightly decreasing trend with higher [Sr/Fe] ratios than those in Draco. \citet{2017ApJ...850L..12T} reported that there was an apparent increase in the abundances of neutron-capture elements at [Fe/H] = $-$2.6 possibly due to the contribution of an NSM. On the other hand, Ursa Minor shows the enrichment of Sr in the entire observed metallicity range  \citep{2010ApJ...719..931C}. In model B$_2$, earlier production of Sr from NSMs makes [Sr/Fe] ratios higher than those in models B and B$_1$. This result implies that an NSM has occurred at [Fe/H] $\lesssim$ $-4$ in Ursa Minor.

\bibliography{sampleNotes}
\end{document}